\documentclass[twocolumn,twoside,slac_two]{revtex4}
\usepackage{graphicx}
\usepackage{fancyhdr}
\begin{document}

\title{Search for Standard Model Higgs in {\boldmath $WH\to \ell\nu b\overline{b}$} at the Tevatron}

\author{Darren D.\ Price (on behalf of the CDF and D\O\ Collaborations)}
\affiliation{Department of Physics, Indiana University, Bloomington, IN 47405, USA}

\begin{abstract}
We present a search for the Standard Model Higgs boson produced in association with a 
W boson in $p\overline{p}$ collisions at a center-of-mass energy of $\sqrt{s} = 1.96$~TeV.  
The search is performed in the $WH\to \ell\nu b\overline{b}$ channel using 2.7~fb$^{-1}$ of data 
collected by both the CDF detector and by the $D\O$ detector, at the Fermilab Tevatron.  
The searches employ artificial neural network, matrix element and boosted decision tree techniques
to improve the separation between signal and background.  Additional 
techniques used to improve the Higgs sensitivity include the use of optimized b-quark 
jet energy corrections and improved algorithms for identifying b-quarks.  
In the absence of an observed excess in data, upper limits are set by both experiments 
on the Higgs production rate times branching ratio for a range of possible Higgs masses between 100 and 150~GeV.
\end{abstract}

\maketitle

\thispagestyle{fancy}

\section{Introduction}

Current experimental results suggest that a Standard Model Higgs boson is likely to be found
at low invariant mass: direct limits the LEP experiments exclude\,\cite{Barate:2003sz} 
a Higgs signal below masses of 114.4~GeV at a 95\% confidence level (C.L.), while more recent 
Tevatron combined results from the CDF and D\O\ experiments exclude\,\cite{TevExclusion:2009}, 
a Higgs between $160-170$~GeV, again at a 95\% C.L.
In addition, electroweak precision fits\,\cite{BlueBandplots} favour a low mass Higgs ($m_H<157$~GeV). 

For low-mass ($115<m_H<150$~GeV) Standard Model (SM) Higgs searches at the Tevatron, 
the dominant production channels are\,\cite{Han:1991ia} (in order of descending cross-section): 
gluon fusion, Higgs production in association with
a $W$ boson, and Higgs production in association with a $Z$ boson. In addition, the SM Higgs
branching fraction at low mass ($m_H<135$~GeV) is dominated by $b\overline{b}$ decays\,\cite{Djouadi:1997yw}. 
The most promising search channel might then be expected to be $gg\to H\to b\overline{b}$, but the QCD process 
$q\overline{q}\to b\overline{b}$ has a cross-section six (or more) orders of magnitude higher than
the expected Higgs production, which all but rules out experimental discovery in this decay mode. 

As such, the most sensitive search channel for a low-mass Higgs at the Tevatron is production
of the Higgs boson in association with a $W$ boson (production in association with a $W$ 
has a rate approximately twice that of production in association with a $Z$), followed by a Higgs decay into $b$-quark pairs
Requiring leptonic decay of the associated $W$ boson provides a distinguishing feature to reduce the large $b\overline{b}$ 
background, making the $WH$ channel one of the most favourable search channels and a significant
component to combined Tevatron search.
Finding evidence for the Higgs in this channel remains extremely challenging as it is rarely produced
($\sigma_{WH}\sim 0.1$~pb) compared to other processes with the same final state, such as $W$ production in
association with jets (which may be true $b$-jets, or misidentified light or charm quark jets), and top
quark production. Because of this, signal-to-background ratios are expected to be very small, on the 
order of $S:B\sim 1:100$. 

Searches for $WH$ production have been recently reported by the CDF\,\cite{CDFWH} and D\O\ \,\cite{DZeroWH} 
collaborations. This paper presents updated results of the $WH\to \ell\nu b\overline{b}$ ($\ell = \mu^\pm, e^\pm$) 
search channel utilising the latest analysis
techniques and using data corresponding to an integrated luminosity of 2.7~fb$^{-1}$ 
collected by the CDF\,\cite{CDF} and D\O\ \,\cite{DZero} detectors at the Fermilab Tevatron
$p\overline{p}$ collider operating at a center-of-mass energy of $\sqrt{s} = 1.96$~TeV.

\section{Event Selection}

The observable final state in this Higgs search channel is two $b$-jets (jets originating from $b$ quarks)
coming from the Higgs decay along with the associated $W$ decay products, a high $p_T$ lepton and the
presence of large missing transverse energy in the event due to the neutrino.

Events are considered as $WH$ candidates only if they have exactly one lepton candidate, with $E_T>20$~GeV
for electrons ($>15$~GeV for D\O\ ) and $p_T>20$~GeV for muons. Leptons from $W$ decays are well isolated from the 
rest of the event, so the cone $\Delta R = \sqrt{\Delta\eta^2+\Delta\phi^2}=0.4$ around the lepton is 
required to contain less than 10\% of the lepton energy. Both jets are required to have an $E_T>20$~GeV and 
$|\eta|<2.0$ ($<2.5$ for D\O\ ) and one or more of the jets is required to have been identified (``tagged'')
as having come from a $b$-jet.

Leptonic $Z$ decays and $t\overline{t}$ dilepton backgrounds could fake a single lepton signature if one 
of the leptons were missed in reconstruction. Both experiments take steps to reduce contamination from
such sources. As an example, CDF reduces fakes from $Z$ decays by rejecting an event if a track, EM cluster or jet 
together with the identified lepton forms an invariant mass between 76 and 106~GeV.

Selected events are required to have a missing transverse energy ($\not\!\!E_T$) greater than 20~GeV (25~GeV
for the electron channel at D\O\ ) in order to be consistent with the presence of a neutrino in the signal $W$ decay. 
Background contamination and multijet processes are higher in the forward region, so CDF applies a
stricter criteria on the $\not\!\!E_T$ (25~GeV) for forward electron events to improve rejection of hadronic backgrounds.
Purity of the sample can be improved by reducing fake events from QCD processes. In the case of events with one $b$-tag,
CDF applies a transverse mass cut on the $W$ ($m_T(W)=\sqrt{2p^\ell_Tp^\nu_T(1-\cos(\phi_\ell-\phi_\nu))}>20$~GeV)
and a cut relating the $\not\!\!E_T$ to the azimuthal angle between the $\not\!\!E_T$ vector and each of the jets 
($\not\!\!E_T>45-30\cdot\Delta\phi$). A requirement for large $\not\!\!E_T$ significance 
(the ratio between the $\not\!\!E_T$ and a weighted sum of factors correlated to mismeasurement, 
such as $\left(\not\!\!E_T,\textrm{jet}\right)$ angles and the size of jet energy corrections) also helps with background rejection.

\section{Signal and Background Modelling}

\begin{figure}[htb]
\centering
\includegraphics[width=60mm]{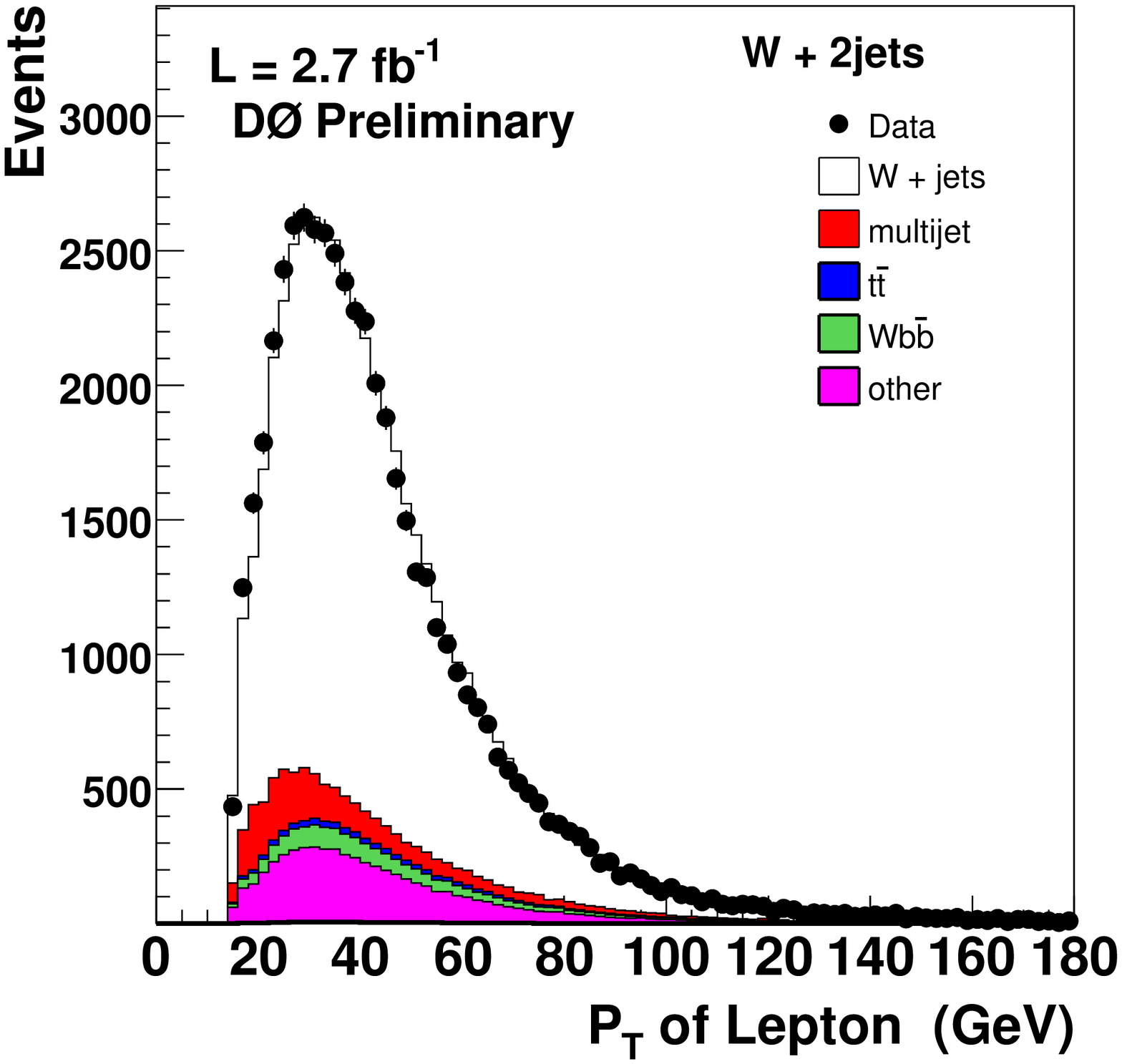}
\includegraphics[width=60mm]{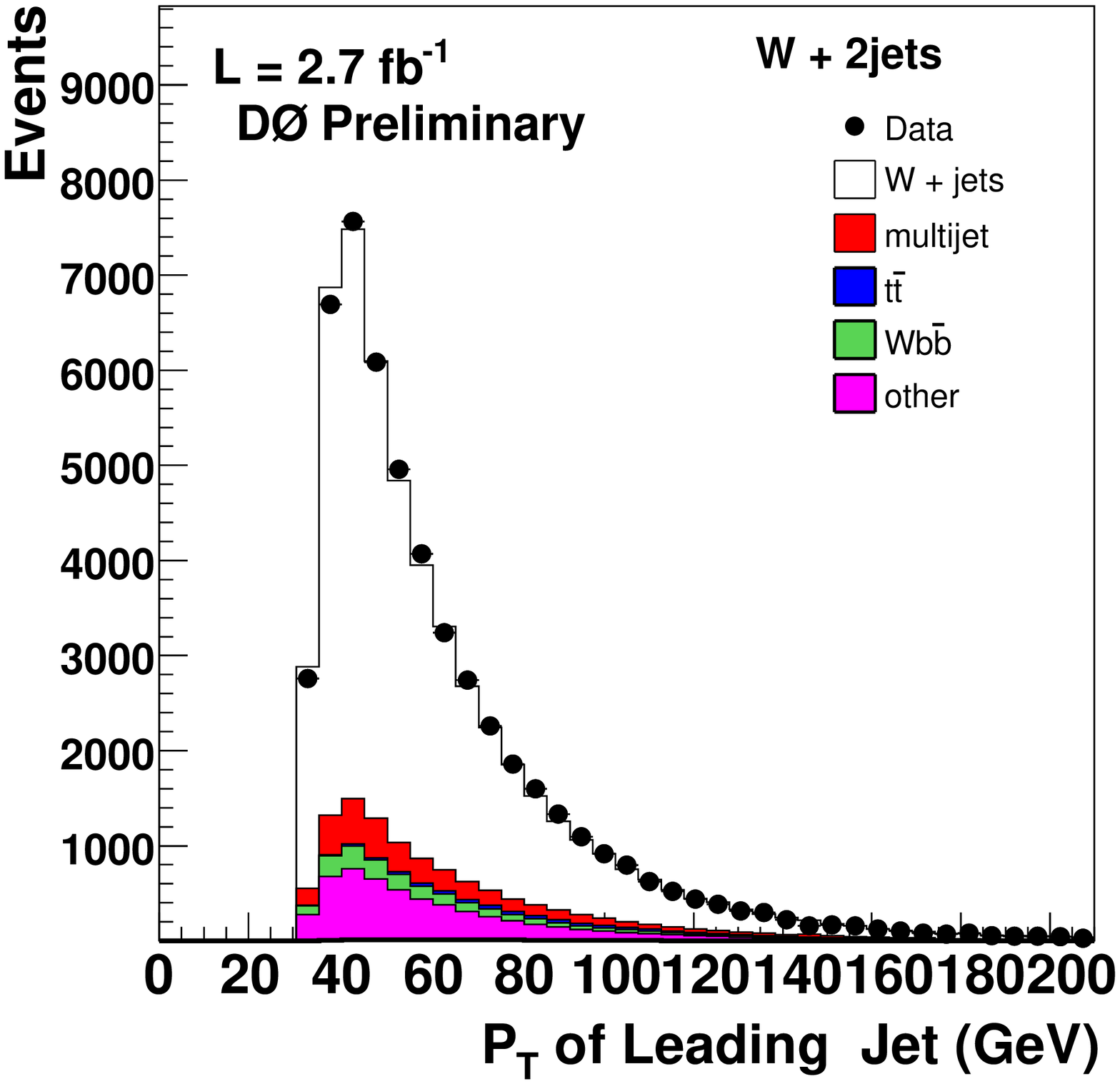}
\caption{Lepton and leading jet $p_T$ distributions in $W$+2 jet data sample (no $b$-tagging) 
compared to simulated expectation. Simulations normalised to integrated luminosity of data sample
and absolute expected cross-sections, with the exception of the $W$+jets sample, which is 
normalised to data in the untagged samples after taking into account all other backgrounds.} \label{fig:H65F01a_H65F02a}
\end{figure}

Figure~\ref{fig:H65F01a_H65F02a} shows examples of the transverse momentum distributions of the lepton and leading (in $p_T$) jet
in the $W$+2 jet data sample before $b$-tagging is applied. 
Good agreement between data and signal and background expectations is seen in these samples, which are used as
a control sample for validation of backgrounds.

The Monte Carlo event generator \textsc{Pythia}\,\cite{Pythia} is used to generate diboson processes (inclusive decays of $WW$,
$WZ$ and $ZZ$), $WH\to \ell\nu b\overline{b}$ and $ZH\to \ell\ell b\overline{b}$ production. \textsc{Alpgen}\,\cite{Alpgen}
interfaced to \textsc{Pythia} for parton showering and hadronisation is used to simulate $W$+jets and $Z$+jets events.
These events were produced using the MLM parton-jet matching prescription\,\cite{Alpgen}. \textsc{Alpgen}-produced 
$W$+jets and $Z$+jets events contain $W{Z}jj$ and $W(Z)cj$ processes, whilst additional heavy-flavour jet
processes $W(Z)b\overline{b}$ and $W(Z)c\overline{c}$ are generated separately in \textsc{Alpgen}. 
Similarly, $t\overline{t}$ (in lepton+jet and di-lepton channels) is also
generated using \textsc{Alpgen}. Single-top events ($s$- and $t$-channel) were generated using \textsc{Comphep}\,\cite{Comphep}
using \textsc{Pythia} for the hadronisation. 

All MC-generated events were processed through the respective detector simulation (based on \textsc{Geant}\,\cite{Geant})
and the same reconstruction software as used for the data analysis. These simulated events are then reweighted to take
into account trigger efficiencies and other ID/reconstruction efficiencies. Simulated backgrounds are then normalised
to their respective Standard Model predictions except for the $W$+jets background which is normalised to data in the
pre-tag sample after taking into account all other physics and instrumental backgrounds, 
where the effect from signal contamination is expected to be negligible. 
Heavy-flavour fractions are calibrated in the one $b$-tagged $W+1$ jet bin (CDF) or $W+2$ jet bin (D\O\ ) 
using data distributions which are sensitive to the flavour composition of the events, allowing heavy
and light-flavour jets to be distinguished. As a result, $W/Z$+ heavy-flavour jets processes
have additional scaling factors applied that are consistent with $K$-factors obtained from MCFM\,\cite{MCFM} that
account for NLO effects over \textsc{Alpgen} modelling.

\subsection{Instrumental and Semi-Leptonic Backgrounds}

Not all backgrounds are determined using Monte Carlo generators. Important other backgrounds to consider
for the analysis are instrumental and semi-leptonic backgrounds (referred to as ``Multijet'' backgrounds 
in this paper) due to fake leptons and fake $\not\!\!E_T$. A jet with a high EM fraction 
can pass electron identification criteria, or a photon may be identified as an electron, giving a fake signal, or
a muon from a semi-leptonic heavy quark decay may be misidentified as being isolated.
In addition, mismeasured missing transverse energy may arise simply from mismeasurements of energy, or from
semi-leptonic decays of heavy quarks. Such $\not\!\!E_T$ mismeasurement is difficult to accurately model
in simulation, and thus the calculation of this background is derived from data samples.

\begin{figure}[htb]
\centering
\includegraphics[width=60mm]{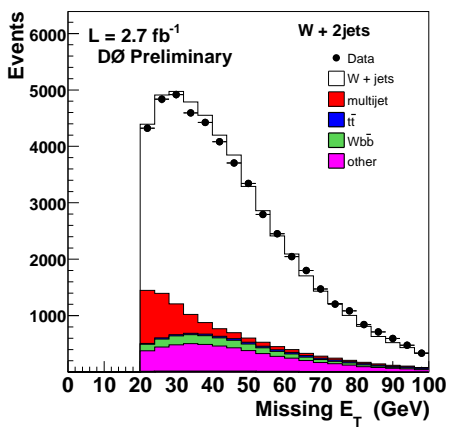}
\caption{Missing transverse energy distribution in the two jet untagged data sample. 
Multijet backgrounds (jets, photons or muons from semi-leptonic heavy quark decays, misidentified 
as electrons or isolated muons) are determined experimentally from an independent data sample and
form a significant background at low $\not\!\!{E}_{T}$.}\label{fig:H65F01d}
\end{figure}

To estimate the number of events containing a jet passing the final electron selection, a sample of data events
are selected with loose lepton requirements, two jets, and low $\not\!\!E_T$. In this
kinematic region fakes dominate the selection criteria and so a probability can be derived (as a function of electron
$p_T$ for example) for a jet faking a loose electron to also pass the tight criteria. A similar technique can
be performed to determine the semi-leptonic background. This multijet contribution receives a further correction,
applied to account for the expected small real lepton contamination of this fake rate, derived from $W$ and $Z$ MC 
simulation in this kinematic region. These probabilities can be extrapolated into the pre-tagged sample selection,
and multijet background distributions built directly from data. The $\not\!\!E_T$ distribution in the pre-tagged
two-jet data sample can be seen in Figure~\ref{fig:H65F01d}, where the multijet component to the other (MC-derived) 
backgrounds can be seen, providing a good description of the data sample.

\section{Tagging of b-jets}

The dominant background to the $WH$ signal in the pre-tagged sample is $W$+jets production, 
due to the overwhelmingly large rate, and the lack of rejection, at this stage, of light-flavour jets.
To extract evidence of a Higgs signal from the recorded events then requires 
excellent identification (``b-tagging'') of real $b$-quark jets in the event and advanced analysis techniques
to accurately model the signal and background contributions to the event selection and discriminate
between them. Once $b$-tagging is applied, the dominant backgrounds come from $Wbb$ and $t\overline{t}$, 
although some light-flavour contamination still remains due to the possibility of ``mis-tagging'' a light jet in a $W$+jet event.
The $b$ quark has a long lifetime (typically on the order of 2~ps), thus $B$ hadrons formed in the hadronisation
of such quarks travel a significant distance (a few millimetres) in the detector before decaying, and thus can be identified by having 
associated tracks displaced from the primary interaction vertex. Making use of this, and other properties of $B$ decays
(such as the higher track multiplicities, and large mass associated with the reconstructed secondary vertex) it is possible
to identify a jet as having originated from a $b$ quark (albeit with some possibility of misidentifying a light quark jet). 

D\O\ uses a neural network algorithm to tag heavy-flavour jets, and assigns tagged events to one of two categories. 
Either both jets in a given event are tagged using a loose selection (which has a mis-tag rate 
of 1.5\% at a jet $p_T$ of 50~GeV), or only one jet is tagged at the loose operating point, in which case a tighter selection 
is used to reduce mis-tag rate to 0.5\%. Thus, we are left with two exclusive samples: a double-tagged selection of two 
loosely-tagged jets, and a single-tagged selection, with one tightly-tagged jet, both of which can be separately
optimised and considered in the Higgs search. The $b$-jet efficiency is $59\pm1\%$ and $48\pm1\%$ for loose and tight selections
at a jet $p_T$ of 50~GeV relative to ``taggable'' jets with particular track quality cuts, which themselves have an typical 
efficiency of $80\%$ in the two-jet bin. 

\begin{figure}[ht]
\centering
\includegraphics[width=60mm]{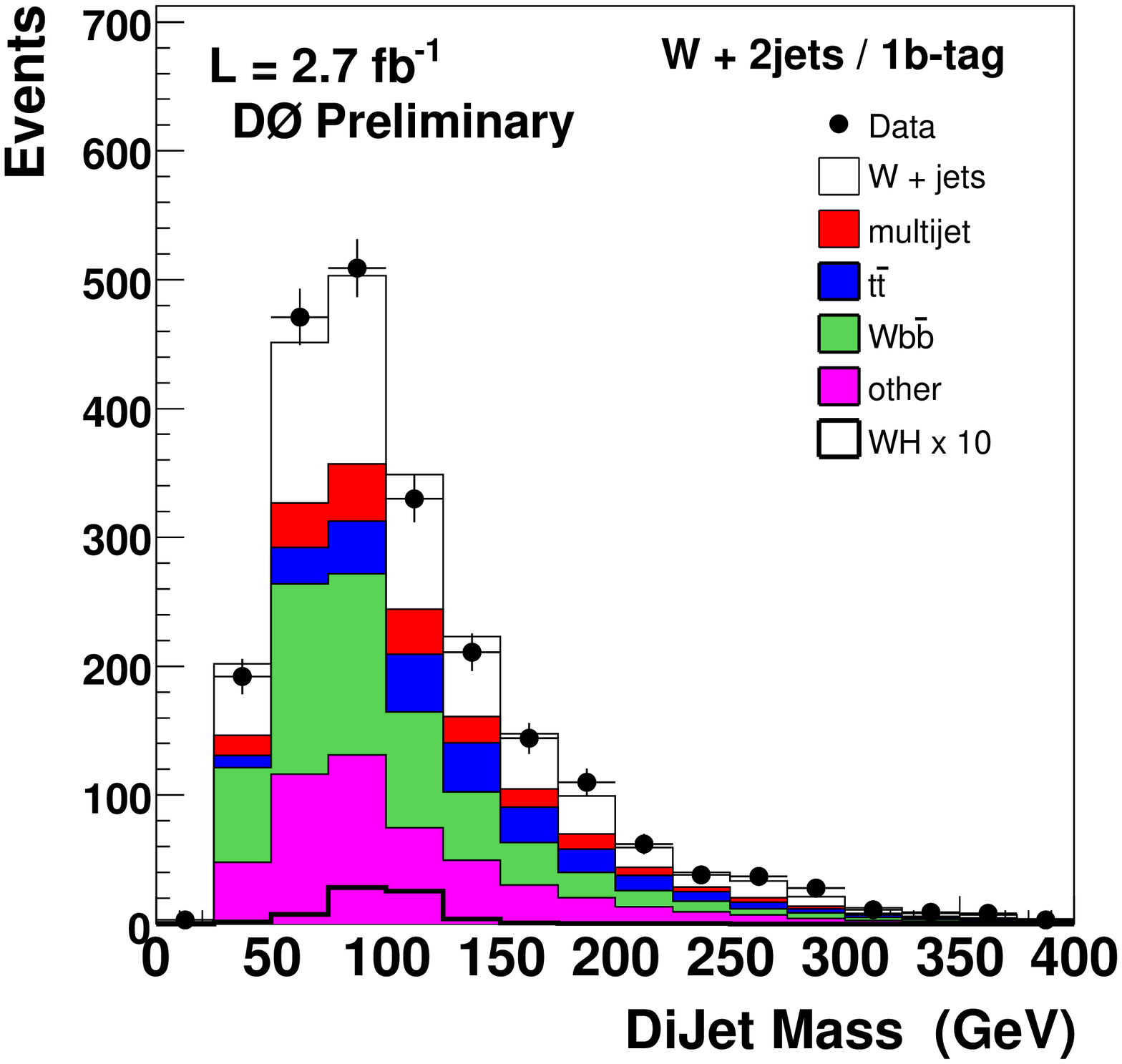}
\includegraphics[width=60mm]{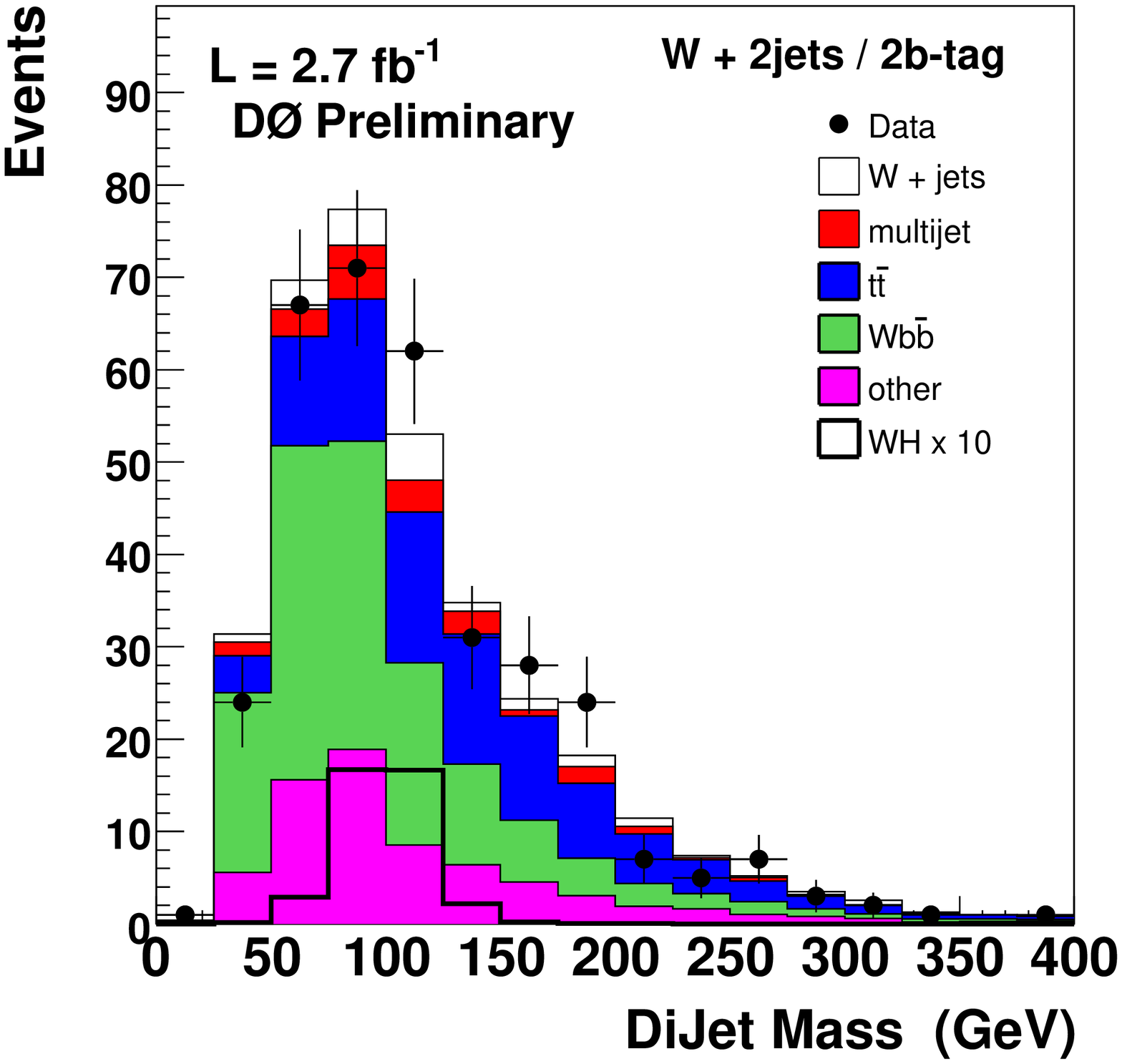}
\caption{Dijet mass distributions in $W$+2 jet events in the single-tagged sample (centre)
and double-tagged sample (right). A pre-tagged sample (see Figure~\ref{fig:H65F01a_H65F02a}) 
is used for normalisation and validation of backgrounds. The orthogonal single and double-tag 
samples are separately considered to improve performance.}\label{fig:H65F05a_H65F05c}
\end{figure}

CDF, in contrast, uses two specialised algorithms and three exclusive tagging categories for $b$-jet identification. 
One algorithm, the \textsc{secvtx} secondary vertex tagging algorithm\,\cite{CDFbtag} identifies $b$-jets by fitting tracks
displaced from the primary vertex, distinguished by their impact parameter significance (ratio of the impact parameter  
to the total uncertainty from tracking and beam position measurements). \textsc{secvtx}-tagged events
are also used as inputs to a neural network, to reduce significant contamination that may result from mis-tags.
The mis-tag rate itself is derived from inclusive jet samples looking
at negative mis-tags associated with an unphysical negative transverse decay length. 
Such a network is trained to separate $b$ from light jets and $b$ from $c$ jets, and uses event information
such as lifetime, invariant mass and track multiplicity as discussed above, in combination with additional
vertex and track parameters to better separate the jets. The neural network is optimised to reject 65\% and 50\%
of light and charm-flavour jets respectively, whilst keeping high (90\%) $b$-jet efficiency.
The second CDF algorithm is the jet-probability tagger, which identifies $b$ quarks by requiring a low probability
for tracks contained within a jet to have originated from the primary vertex, using signed impact parameters on
all tracks within a jet for discrimination. The misidentification rate 
can be calculated from tracks with a negative impact parameter (associated with a negative probability) in the pre-tagged sample.
A jet is considered tagged if it has a jet probability of less than 5\%, which allows 95\% rejection of light
jets with a $b$-jet efficiency of 60\%. The two algorithms are used to produce three exclusive tag samples for the search:
a single-\textsc{secvtx} tag only sample, a double-\textsc{secvtx} tag sample, and a single-\textsc{secvtx} plus one jet-probability tagged
sample (``ST+JP'').

The effect of applying $b$-tagging to the untagged sample is shown by the example in Figure~\ref{fig:H65F05a_H65F05c}, where the
composition of the single-tag and double-tag selections in the dijet mass distribution after tagging is seen.
Light-flavour jets are heavily suppressed, and a possible signal is much enhanced relative to the backgrounds compared
to the untagged case.

\section{Analysis Technique}

The analysis results described here from D\O\ use a combined neural network (NN) and matrix element (ME) based approach,
that make use of the full event kinematics rather than relying on signal discrimination within one
particular kinematic distribution, whilst the CDF results combine two separate analyses: one based on a neural
network approach, and one based on a combined matrix element and boosted decision tree (BDT) method. These two CDF
multivariate analyses are then used as inputs to a further NN super-discriminant which uses the uncorrelated
information between the two analyses to produce a combination limit with improved sensitivity.

\subsection{Matrix Element Discriminants}

The matrix element method used by D\O\ and CDF as further inputs to a neural net and BDT respectively 
builds an event probability density for signal and background processes using the 4-vectors of the lepton and two jets.
If the detectors were ideal, one would be able to define an event probability as a differential cross-section that takes
into account the Lorentz invariant matrix element, incident particle four-momenta and $n$-body phase space, normalised to
the total cross-section. However, effects such as energy resolution and problems identifying the final state neutrino mean
we need to integrate over the unmeasured neutrino momentum and convolve the results a resolution function from the detector
mapping all possible particle-level variables to their observed counterparts. The leading order matrix element itself
is calculated using the \textsc{helas}\,\cite{HELAS} package.

\begin{figure}[ht]
\centering
\includegraphics[width=60mm]{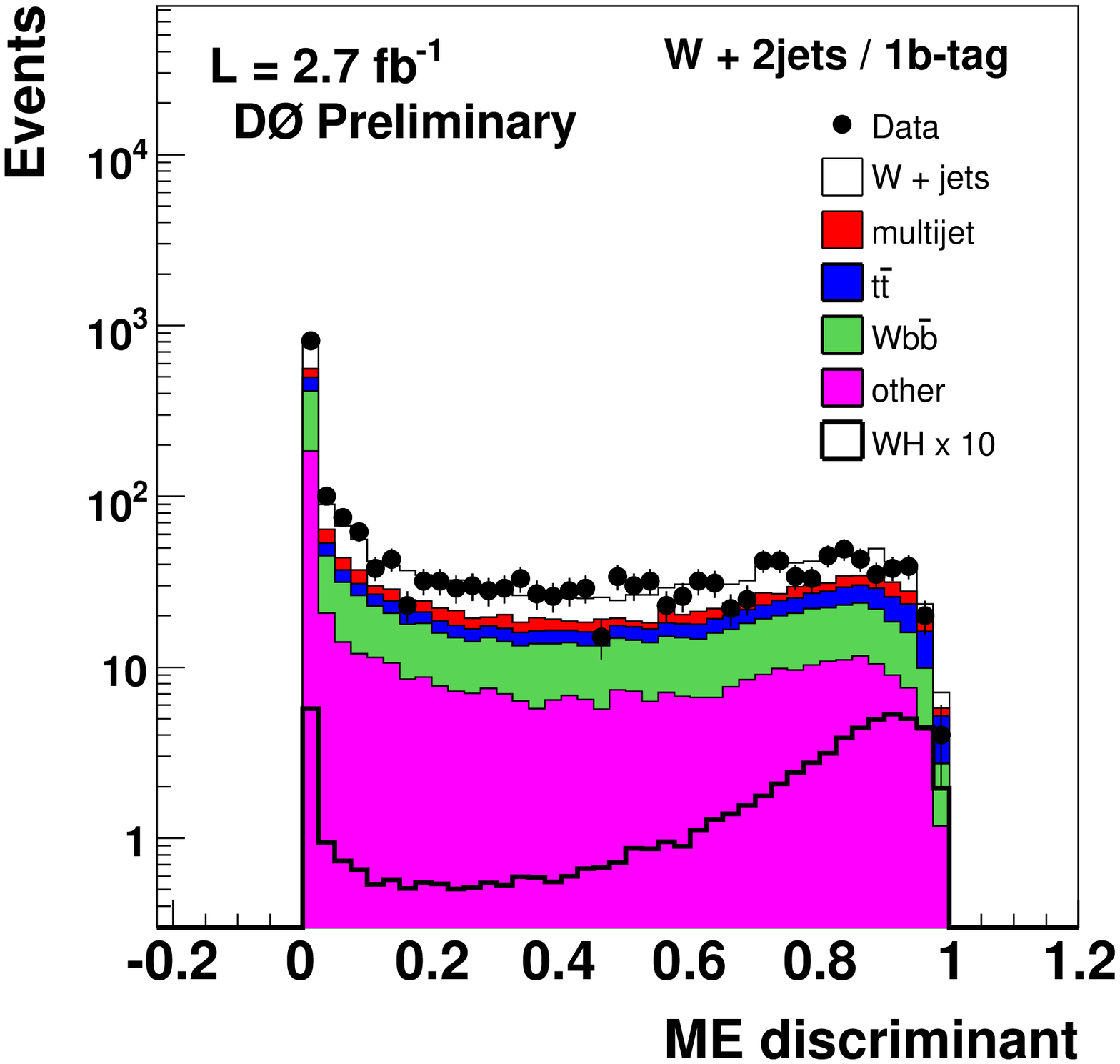}
\includegraphics[width=60mm]{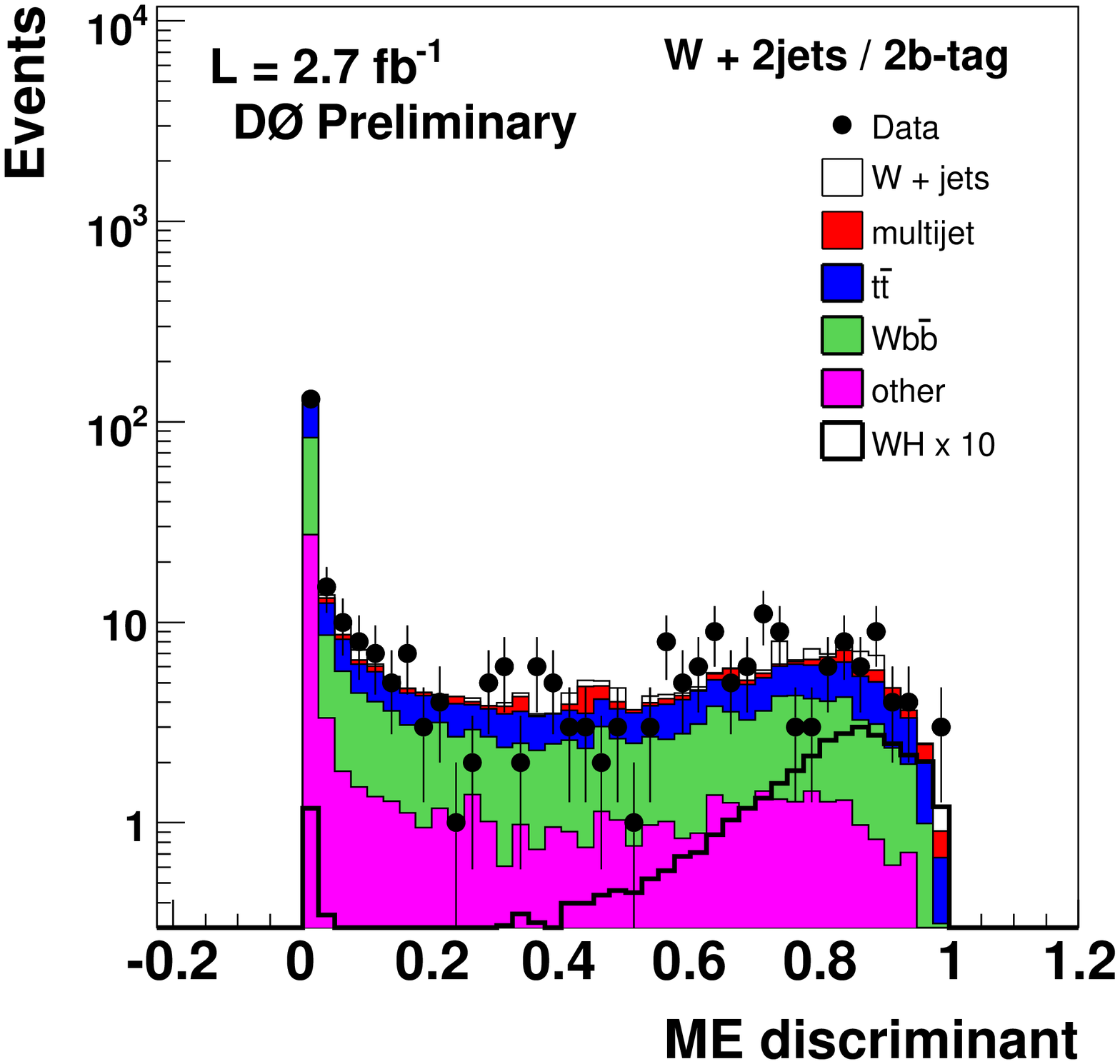}
\caption{Matrix Element discriminant output data distributions compared with expectation in the two-jet one-tag (top)
and two-jet two-tag (bottom) samples. The $WH$ expected signal is scaled by a factor of ten for visibility.} 
\label{fig:H65F07c_H65F07d}
\end{figure}

Once the effects mentioned above are incorporated, an event probability is formed and it is possible to 
calculate the relative probability for a given event to come from a $WH$ decay or from one of the backgrounds. 
An example of the output of the matrix element discriminant method output from D\O\ is shown in Figure~\ref{fig:H65F07c_H65F07d},
where application of the ME method results in the $WH$ expected signal peaking towards higher values of the discriminant.

\subsection{Boosted Decision Trees}

CDF uses their matrix element discriminant is an input to a Boosted Decision Tree (along with the neural network
trained to classify jet flavours). The BDT is a binary tree classifier which performs decisions on a single variable
at a time until some stop criterion is reached. The phase space of the final decision nodes is split into a signal and a 
background classification dependent on how the training samples behaved under the same decisions. ``Boosting'' comes
from the combination of several decision trees, derived from the same training sample, with reweighted events which are
then used to form a ``majority vote'' classifier which enhances the stability of the response of the BDT to fluctuations
in the training sample.

\begin{figure*}[ht]
\centering
\includegraphics[width=80mm]{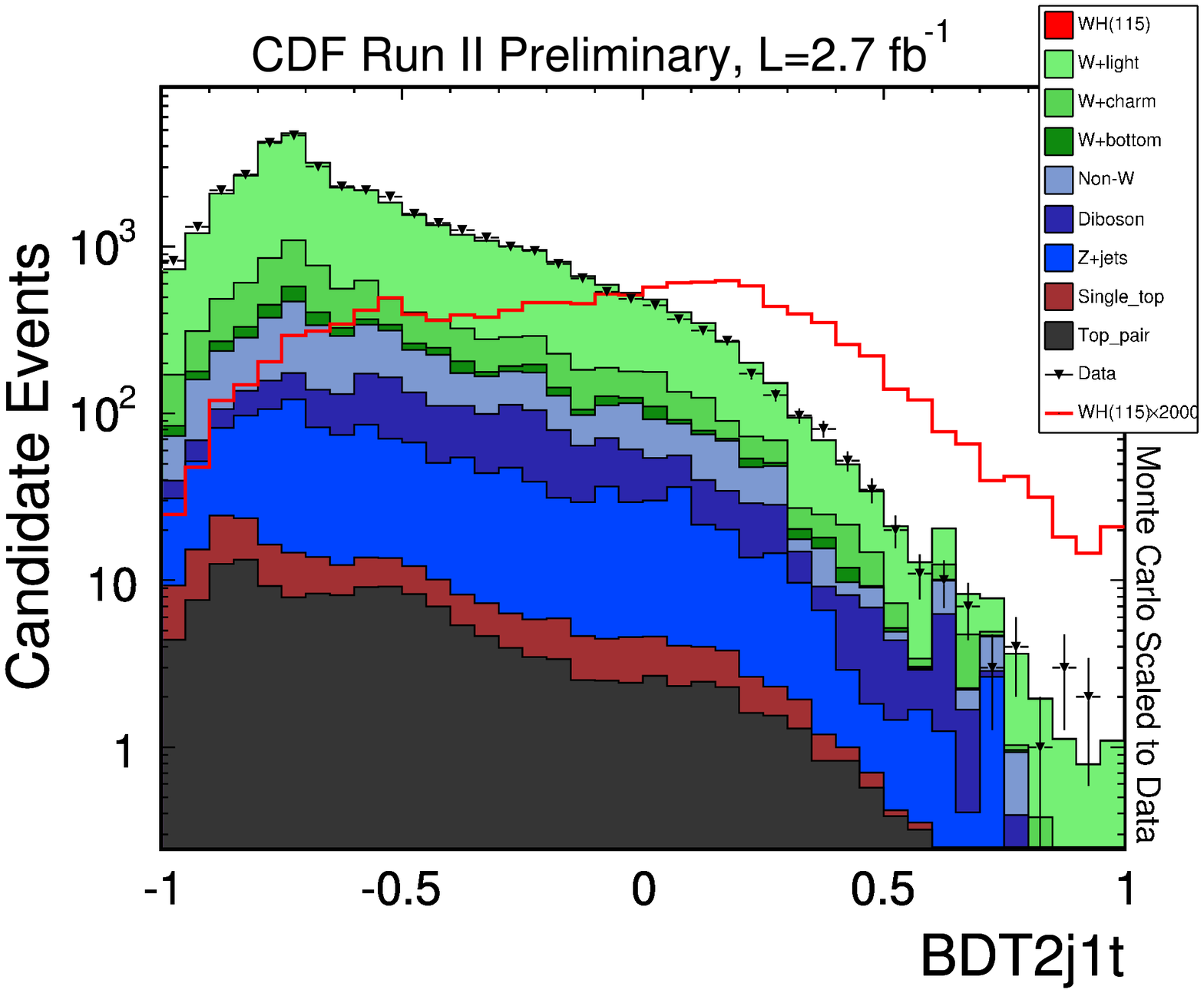}
\includegraphics[width=80mm]{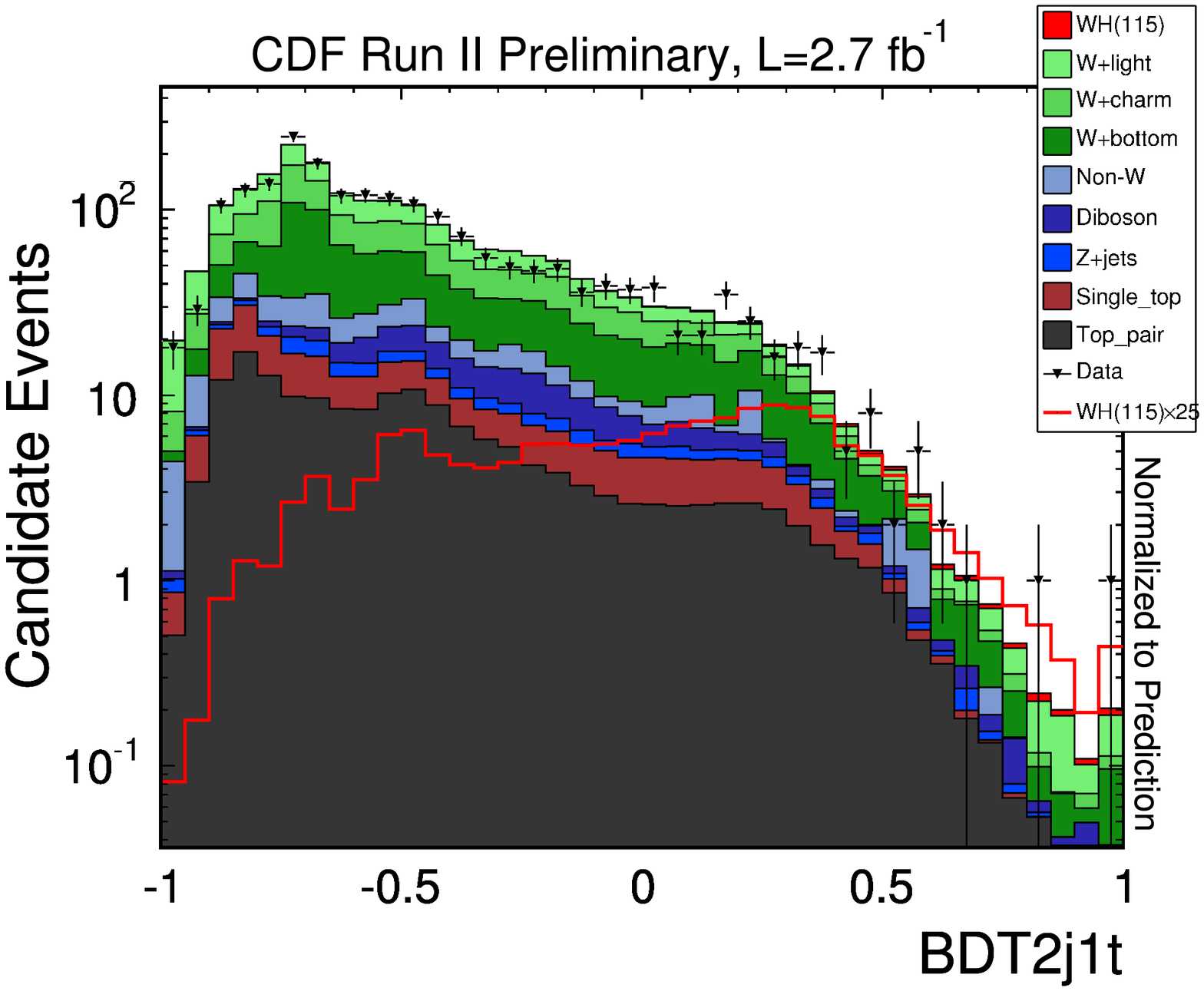}
\includegraphics[width=80mm]{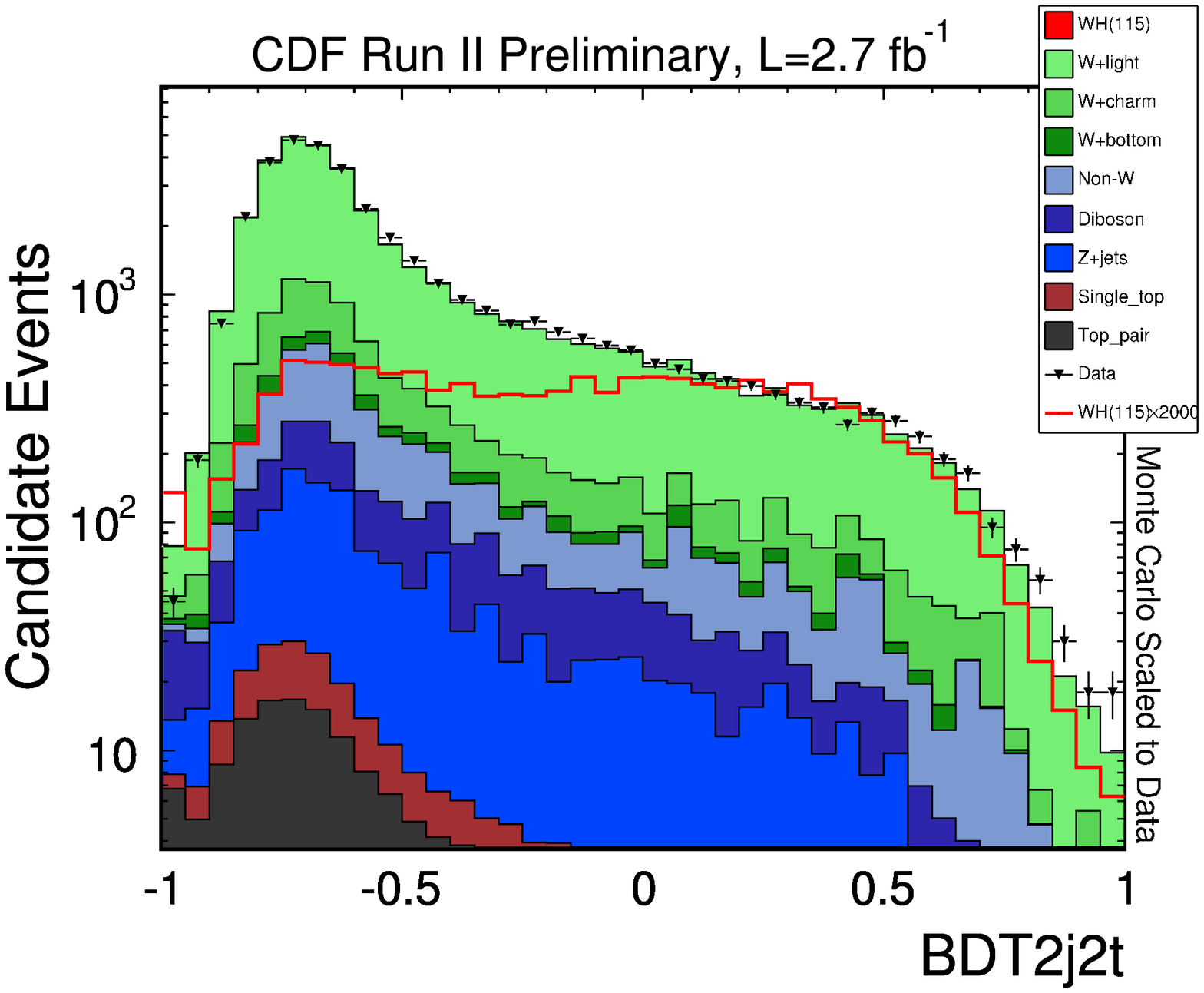}
\includegraphics[width=80mm]{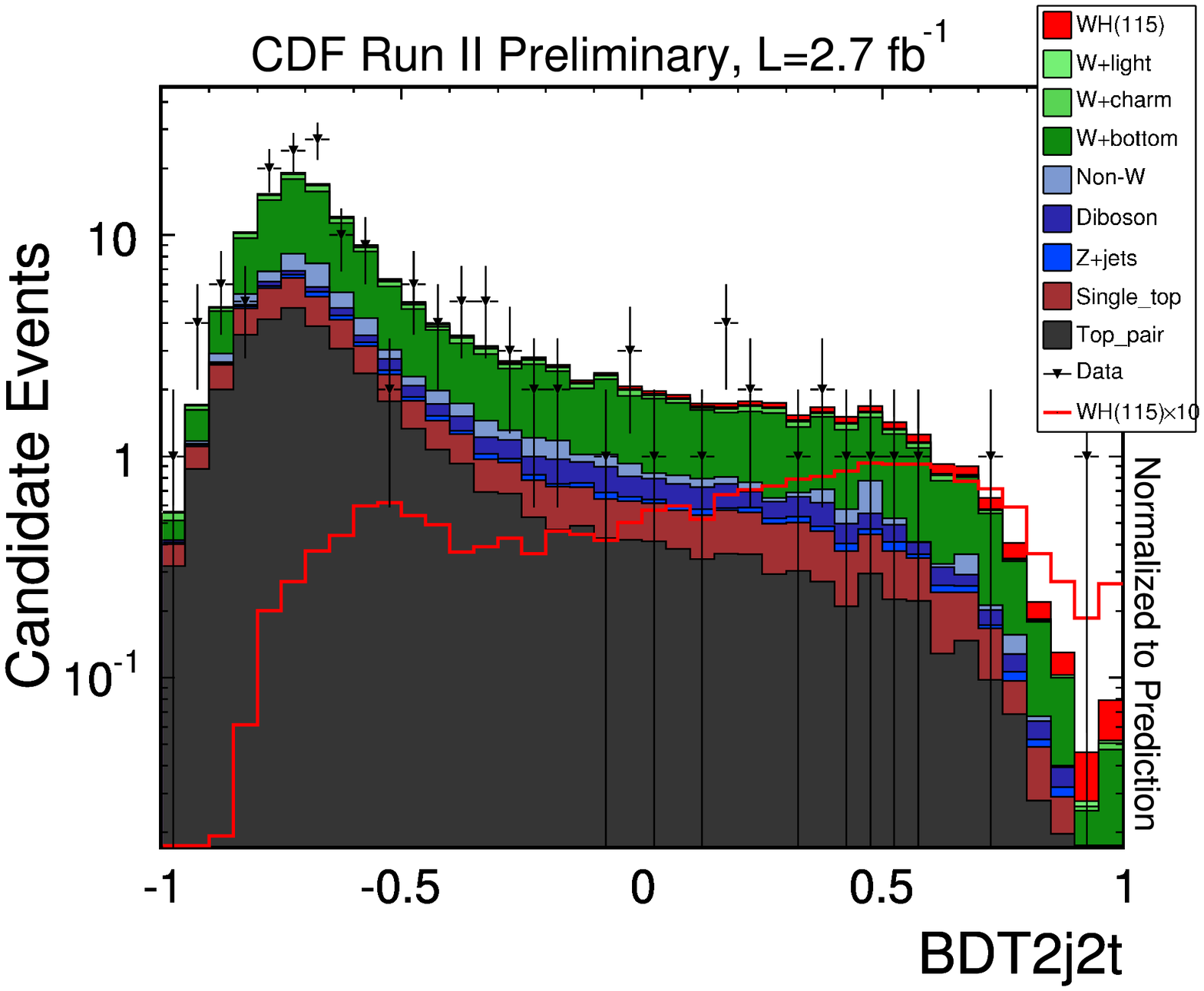}
\caption{Boosted Decision Tree output in the control region with no $b$-tags (left) with a
BDT optimised for the $W$+2 jets one-tag (top-left) and two-tag (bottom-left) signal region, and for 
the signal-enhanced tagged data samples (right) in single-tagged (top-right) and double-tagged 
(bottom-right) data events.} \label{fig:cdf_bdt_out}
\end{figure*}

In addition to the ME and NN discriminants described above, dijet mass, jet $E_T$, event $\not\!\!{E}_{T}$, 
lepton $p_T$ and $\eta$, $\Delta\phi(\textrm{jet}_1,\not\!\!{E}_{T})$, $\Delta\phi(\ell,\not\!\!{E}_{T})$, $m_T(W)$, 
$\cos\theta(\textrm{jet}_1,\ell)$ and the scalar sum of the transverse energies in the event are all combined 
into separate BDTs for the one-tag and two-tag samples, with BDTs trained for specific Higgs mass values. 
The nature of the BDT approach means that poorly discriminating variables do
not affect the final discrimination, as the training algorithm will effectively remove such variables from consideration.
The control region (two-jet pre-tag) is used for validation of the BDT behaviour (shown on the left of 
Figure~\ref{fig:cdf_bdt_out}). The BDTs are then applied to the one-tag and two-tag data samples, and the discriminating
power of this approach is shown in Figure~\ref{fig:cdf_bdt_out} (right), illustrated for a Higgs signal at a mass of 115~GeV
peaking toward higher values of the BDT output.

\subsection{Neural Network Discriminants}

Both CDF and D\O\ use neural network based analyses for the $WH$ search, that utilise the differences between 
kinematic properties of objects in the event to separate a Higgs signal from background events. 
The D\O\ NN uses their matrix element discriminant method along with six other variables (jet $p_T$,
$\Delta R(\textrm{jets})$, $\Delta\phi(\textrm{jets})$, $p_T$(dijet system), dijet mass, and the $p_T$ of the 
$(\ell-\not\!\!{E}_{T})$ system) as inputs to training samples in $WH$ and $Wbb$ simulated events for each of
the electron and muon, single-tag and double-tag, and Run IIa and Run IIb channels (eight separate networks).
The most discriminant kinematic variable is the dijet invariant mass, and this is used as the final discriminating
variable for the $W+3$~jet sample. 

\begin{figure*}[ht]
\centering
\includegraphics[width=52mm]{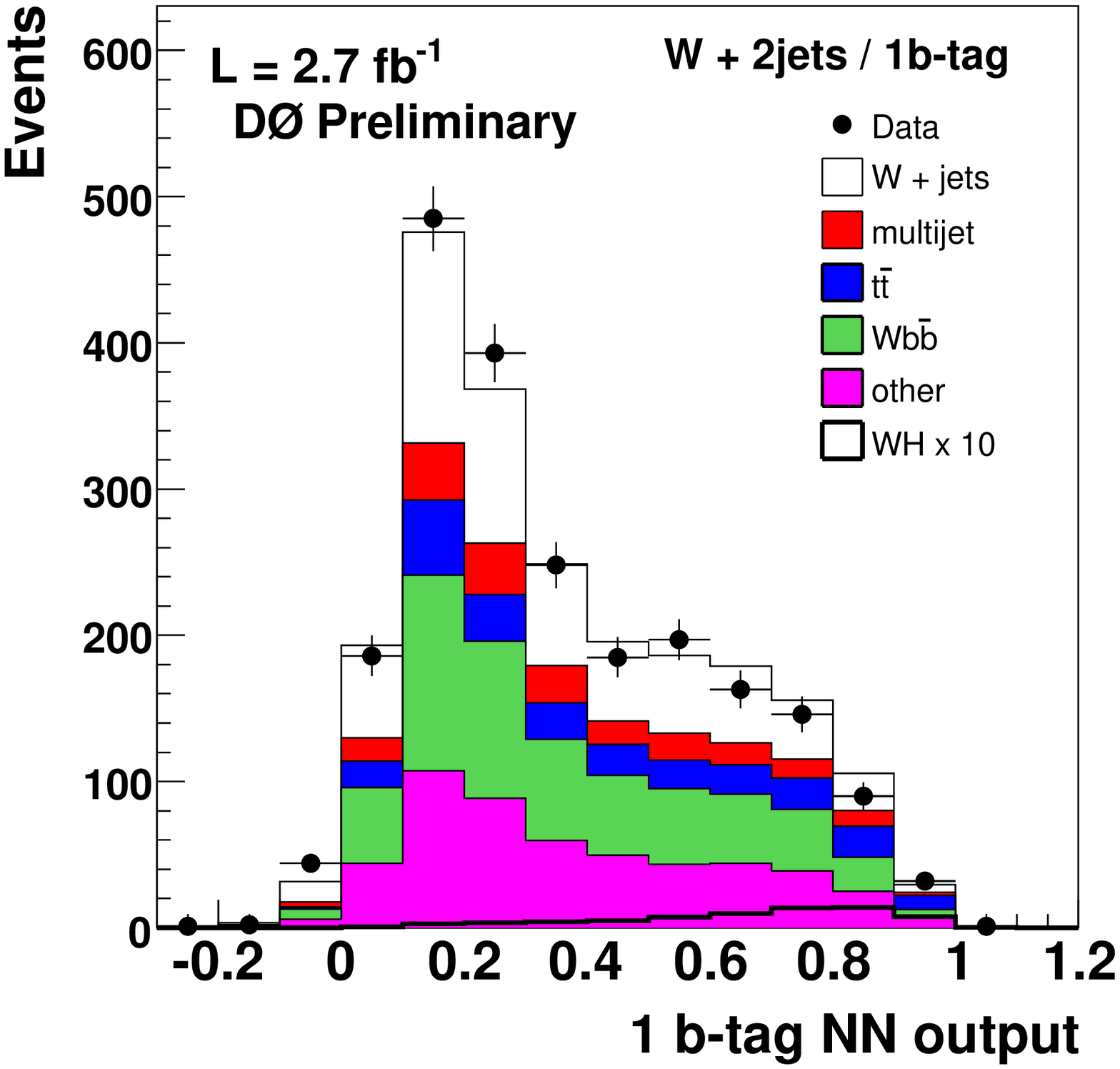}
\includegraphics[width=52mm]{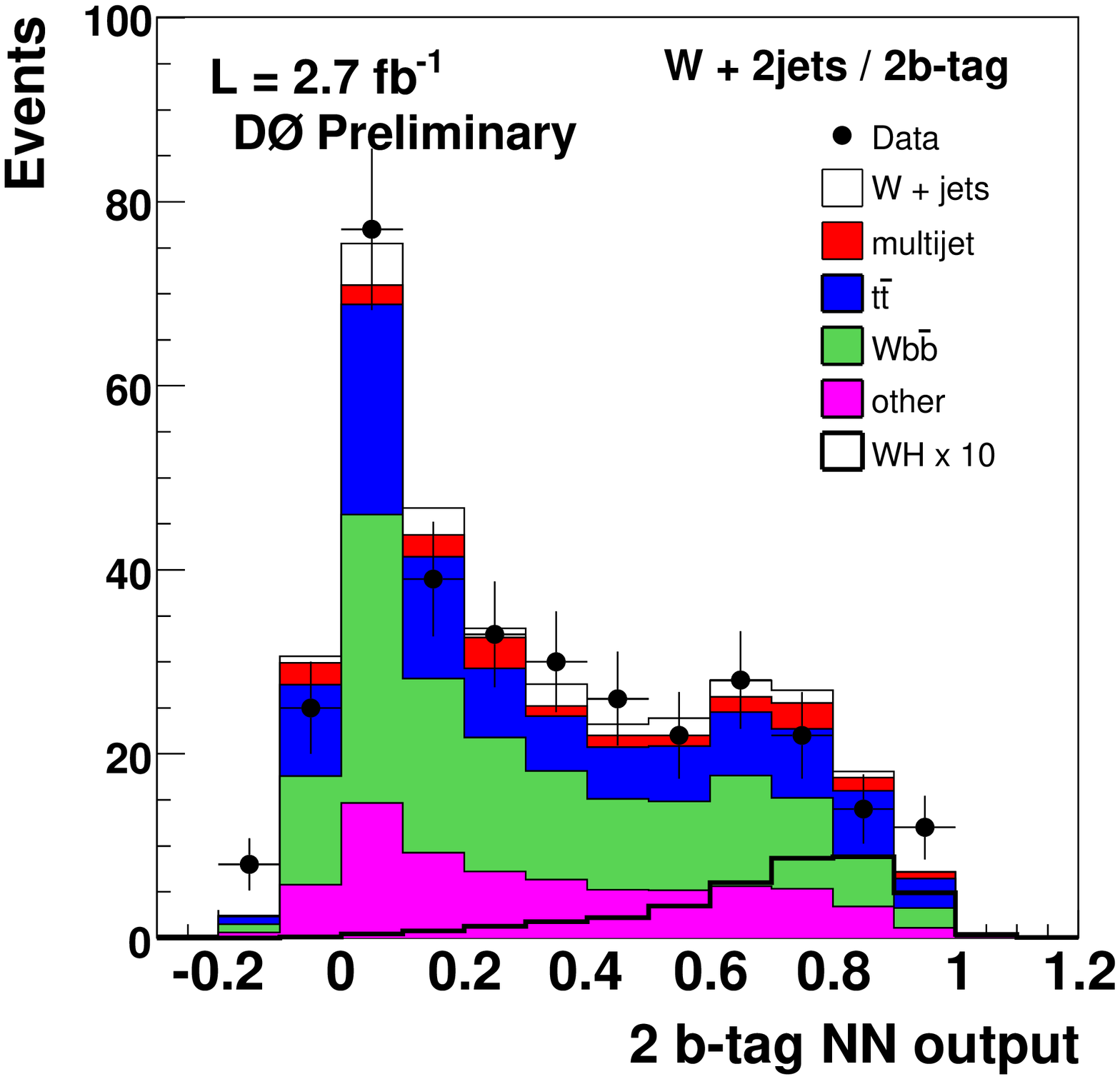}
\includegraphics[width=64mm]{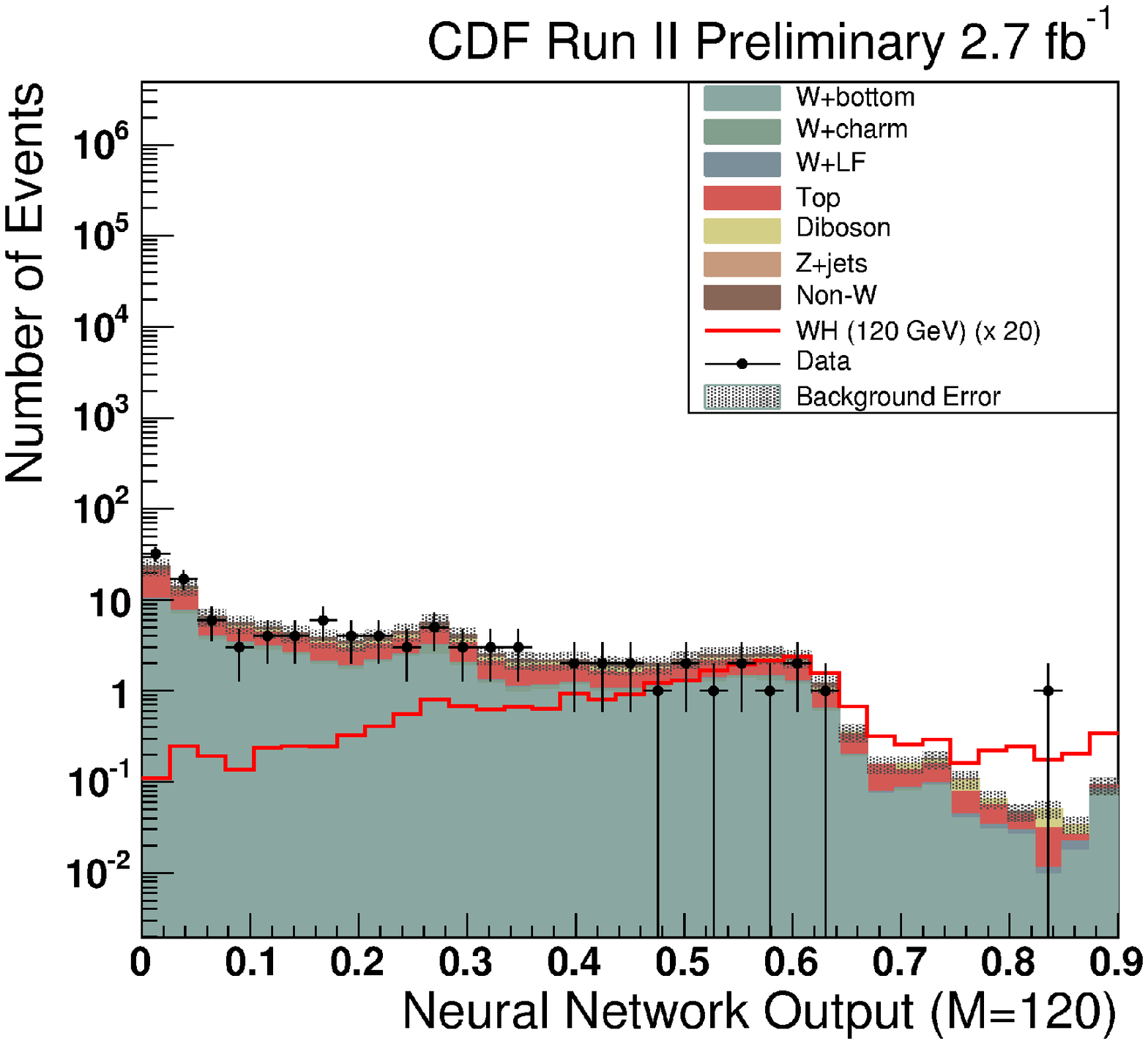}
\caption{Distributions of neural network output compared with simulated expectation in the single-tag sample 
for the single-tag NN (left), the double-tag sample for the double-tag NN (centre), and the CDF ST+JP tagged sample
(right). The $WH$ signal is scaled by a factor of 10 (20 for CDF ST+JP) for visibility. 
} 
\label{fig:H65F08c_H65F08d_STJPcdf}
\end{figure*}

The matrix element discriminant used as an additional input to the
neural network gives an additional 5\% sensitivity in the results over using kinematic distribution
inputs alone. The results of the NN discriminant from D\O\ for the two-jet bin is
shown in Figure~\ref{fig:H65F08c_H65F08d_STJPcdf} for the single and double-tag bins, where the signal clearly
peaks at high values against the background. The expected sensitivity gain from using the combined ME+NN approach over just the
dijet mass as a discriminant is of the order of 20\% (dependent on Higgs mass), which is the equivalent of
a 40\% increase in integrated luminosity. 

CDF has a dedicated NN analysis which uses six inputs chosen from a list of 76 possible variables, iteratively chosen by an optimisation
procedure that looks at the effect of a particular variable on the NN output. 
The six variables chosen are: the invariant mass of the two jets (plus additional `loose' jets if
they lie within $\Delta R=0.9$ of a primary jet); vector sum of the $p_T$ of the lepton, $\not\!\!{E}_{T}$ and two jets;
$p_T$ imbalance (the scalar sum of lepton plus jet $p_T$ minus $\not\!\!{E}_{T}$); scalar sum of loose jet $E_T$;
$\Delta R(\ell-\nu)$ (where the $p_z$ of the neutrino is chosen to be the largest $|p_z|$ from the $W$ transverse mass
constraint); and $m(\ell,\nu,j)_{min}$, where the invariant mass is minimised by choice of the primary jet to include.
The neural network is trained on a mixture of signal and background samples and optimised to place signal-like events
at high NN output, and the NN distributions checked in the pre-tag and tagged samples. An example of the NN output from CDF
is shown in Figure~\ref{fig:H65F08c_H65F08d_STJPcdf} (right) for the ST+JP sample.

\begin{figure*}[htb]
\centering
\includegraphics[width=53mm]{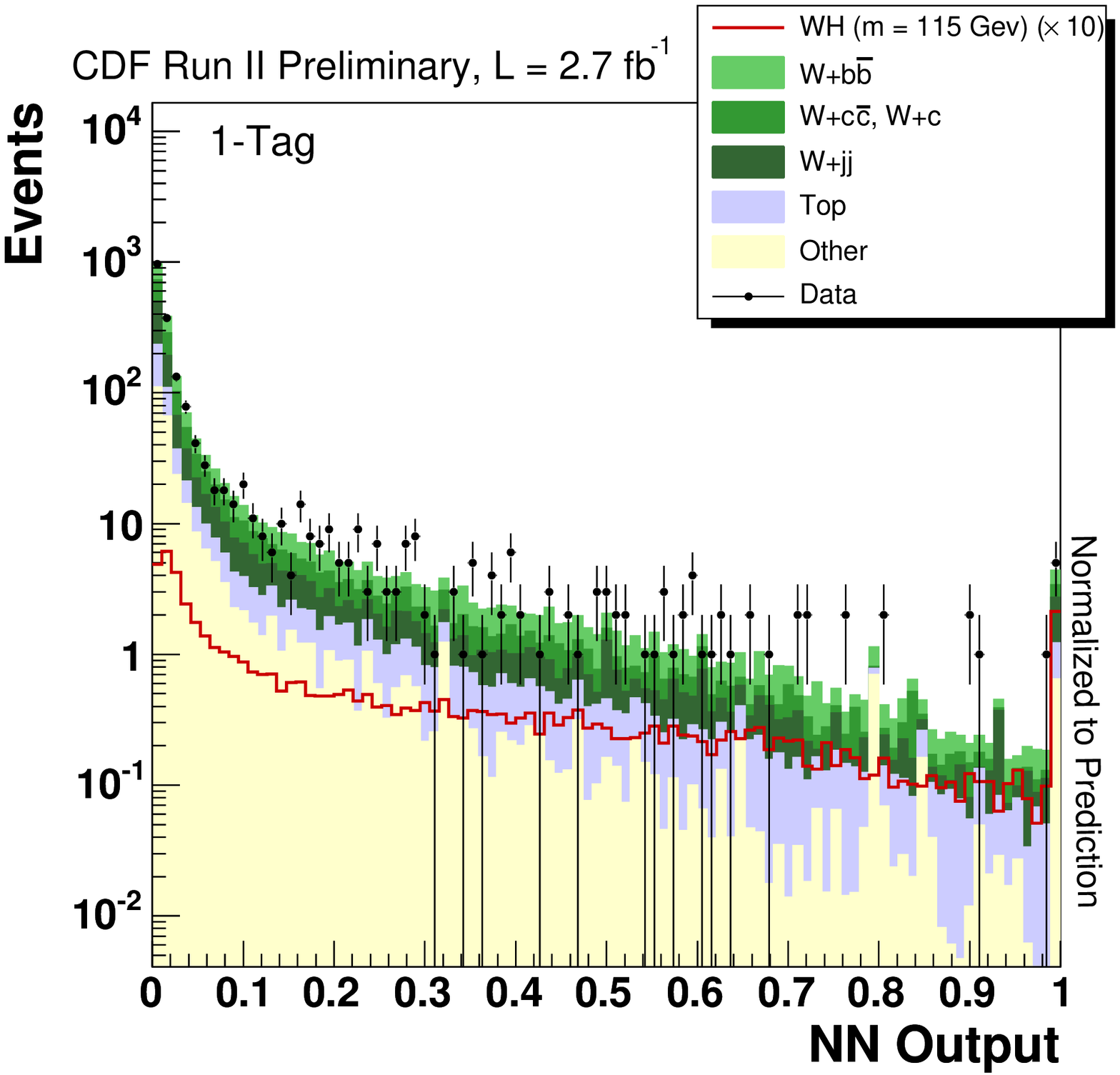}
\includegraphics[width=53mm]{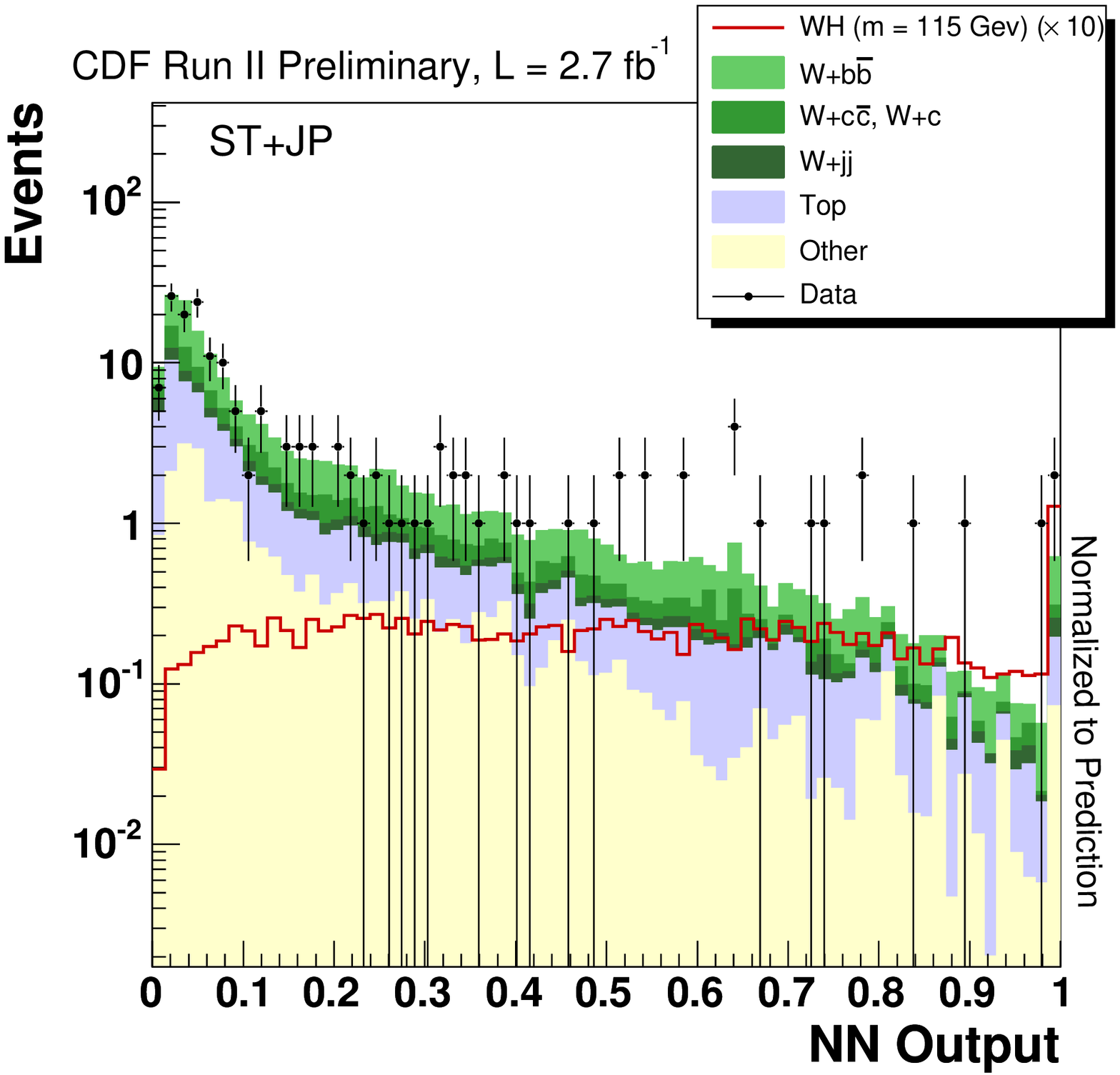}
\includegraphics[width=53mm]{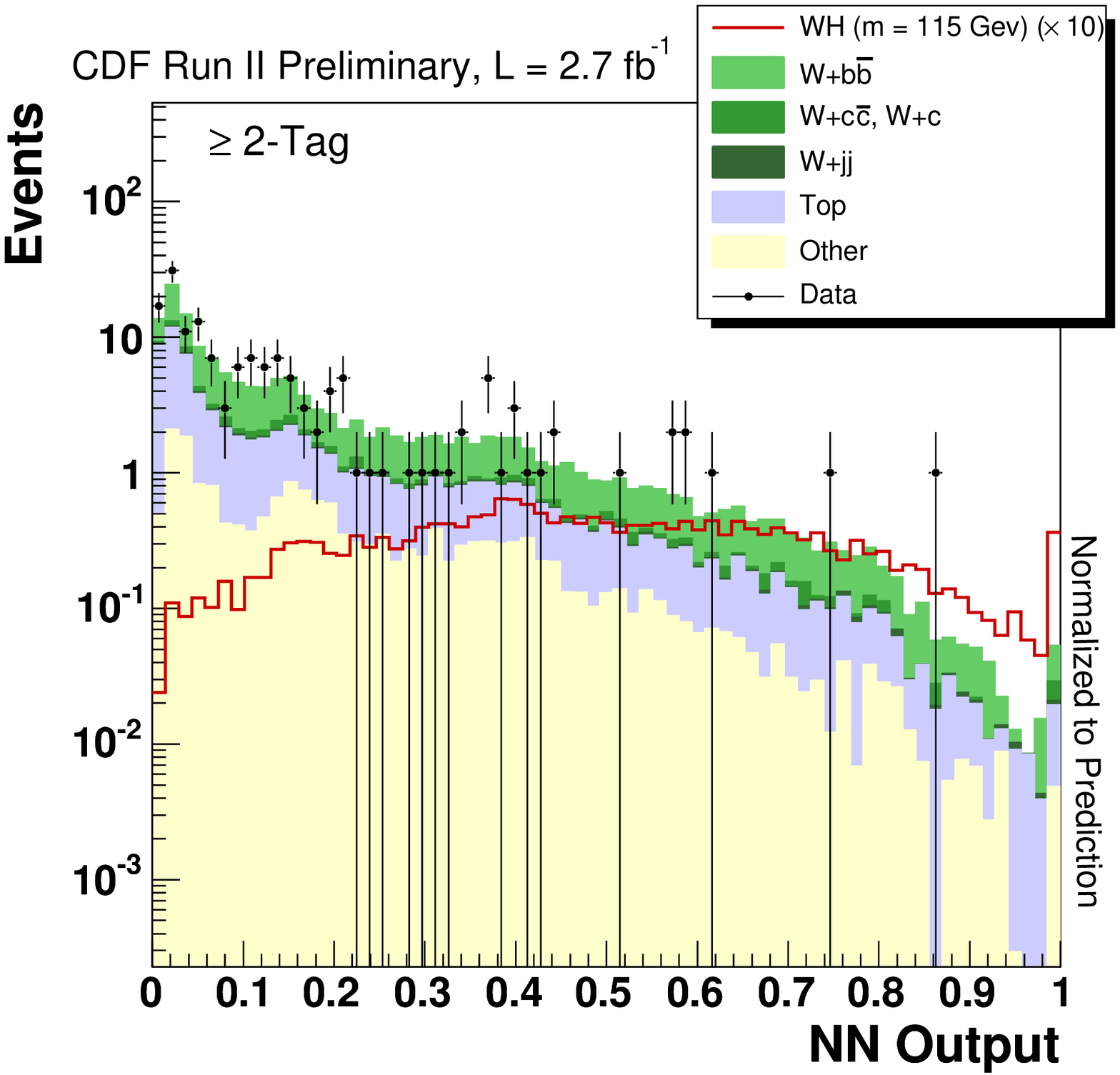}
\caption{CDF NEAT output distributions from the combined NN and ME+BDT analysis inputs for signal and background, in the one-tag (left),
two-tag (centre) and ST+JP tag (right) bins.} \label{fig:cdf_neat}
\end{figure*}
In addition to the independent NN analysis, CDF also performs a further NN analysis combining the results
from the NN and ME+BDT methods as inputs to a super-discriminant technique, first used in the single-top search\,\cite{SingleTop}.
Discriminants from the two analyses are correlated at the level of $50-75\%$, but this leaves room for
sensitivity gain in combining the two methods. CDF makes use of genetic algorithms to stochastically optimise the NN 
architecture and node weights to provide the greatest sensitivity for the search. The package ``Neuro-Evolution of 
Augmenting Topologies'' (NEAT)\,\cite{NEAT} is used to perform this optimisation. This allows additional complexity
to be added to the networks as needed to improve performance. Training is performed with half of available signal
and background MC samples, with the other half reserved for final checks of over-training. As a figure of merit for the
NEAT network fitness to optimise, the expected signal over square root of the expected background is used (as using the full
expected limit is computationally expensive). The NEAT algorithm produces networks trained separately for the single-tag,
double-tag and ST+JP data samples, the outputs of which are shown in Figure~\ref{fig:cdf_neat} for a Higgs mass of 115~GeV
(the training procedure is repeated at a range of Higgs masses). The super-discriminant analysis improves sensitivity 
in the Higgs mass range studied by $5-15\%$ compared to the the individual CDF analyses alone.

\section{Systematic Uncertainties}

Systematic uncertainties on the signal acceptance and the shape of the discriminant distributions can have significant
effects on the Higgs sensitivity. Uncertainties due to lepton ID and reconstruction/trigger efficiencies can be determined
by propagating their effect on acceptance in MC samples.
 
D\O\ assigns a 3-5\% uncertainty for the trigger efficiency used in the analysis,
and a 5-6\% uncertainty on the lepton selection (CDF assigns 2\% uncertainty). CDF calculates uncertainties on the jet energy scale 
corrections to be 2\%, the effect of initial and final state radiation (by halving and doubling the initial MC parameter default values) 
and PDF uncertainties to be 3.1-5.6\% (combined), and systematics from $b$-tagging to be 3.5-8.4\%. 
D\O\ assigns a 2-6\% uncertainty on acceptance from jet ID and energy calibration/resolution uncertainty, and 5\% due to jet modelling.
For D\O\ $b$-tagging uncertainties come from jet taggability (3\%) and $b$-tagging efficiency (2-5\%), per heavy-quark jet. Uncertainties
on light-quark jets translate into an uncertainty on the total background of 7\% in the single-tag sample, and are negligible in the
double-tag. An uncertainty ($\sim$10\%) is assigned on the NN output to account for $W+$jets modelling uncertainties, derived by comparing
the original distribution to one reweighted to data in the pre-tag sample. An additional 5-10\% uncertainty is added for $Wbb$ invariant
mass shape modelling to account for differences in data to MC in the pre-tag sample. Overall, the experimental uncertainty on the acceptance
calculated by D\O\ varies between 16\% and 28\% dependent on both process and channel (and as a guide is $\sim$18\% for $WH$ in the 
double-tag sample), with the uncertainty on background cross-sections being 11\% for $t\overline{t}$ and single top, 
6\% for diboson production and 20\% for W+ heavy-flavour production.

\section{Results}

\begin{figure*}[tb]
\centering
\includegraphics[width=80mm]{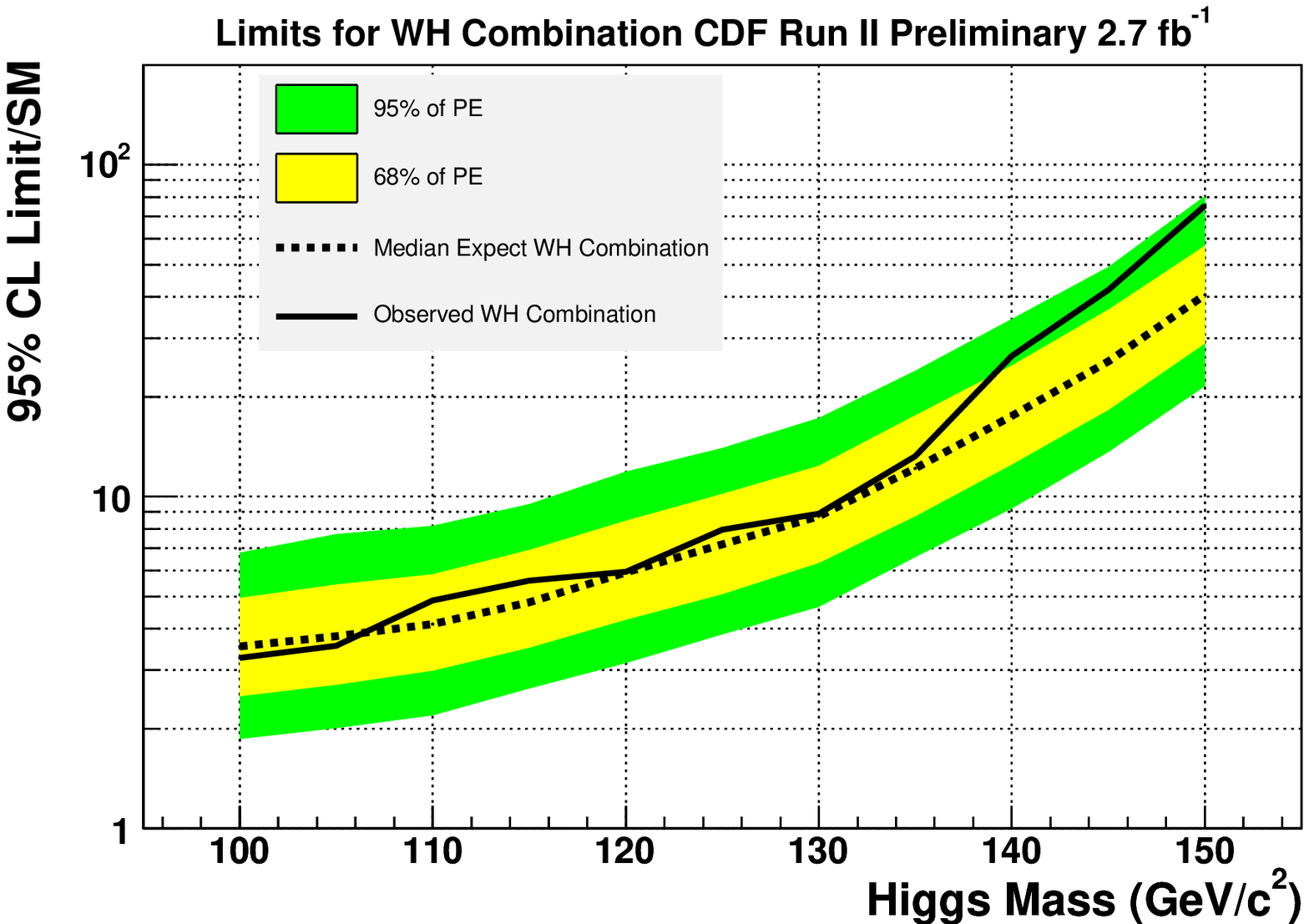}
\includegraphics[width=85mm]{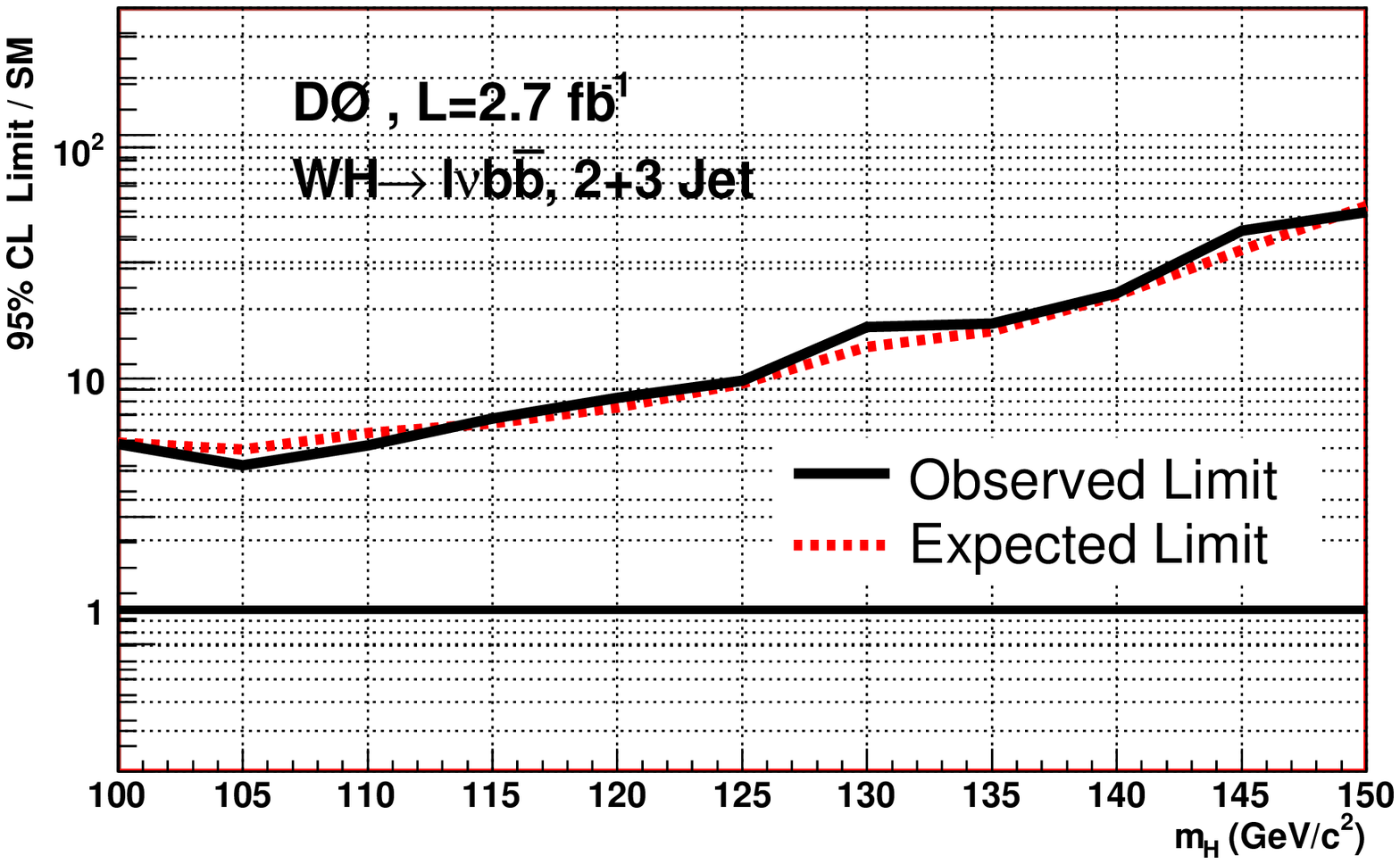}
\caption{CDF (left) and D\O\ (right) expected and observed 95\% confidence-level upper limits on cross-section ratios to the 
Standard Model expectation, for the combined $WH\to \ell\nu b\overline{b}$ analyses from each experiment for a range of possible Higgs masses 
(one and two sigma bands within which the limits may fluctuate, in the absence of signal, are shown on the expected upper limit plot from CDF). } 
\label{fig:limit_ratio}
\end{figure*}

Searches for the Standard Model (SM) Higgs boson in the $WH\to \ell\nu b\overline{b}$ channel were performed using neural network,
boosted decision tree and matrix element approaches from D\O\ and CDF.
No evidence of a Higgs boson is observed and so upper limits are placed on its production rate for a range of Higgs masses,
calculated at the 95\% confidence level for the CDF NEAT and NN+ME D\O\ results. 
The impact of all systematic uncertainties is taken into account, as are correlations between channels and between signal and backgrounds.
These limits (as a ratio to the SM cross-section) are listed in Table~\ref{tab:limit_table}
and displayed in Figure~\ref{fig:limit_ratio} for CDF (left) and D\O\ (right), and represent a significant gain
in sensitivity over previous searches coming in part from a larger dataset but also extended signal acceptance,
improved $b$-tagging and new multivariate techniques. The combined results at $m_H=115$~GeV are $5.2\times\sigma_{SM}$ observed
($5.3$ expected) at D\O\ and $3.3\times\sigma_{SM}$ observed ($3.5$ expected) at CDF.

\begin{table*}[htb]
\begin{center}
\caption{Expected and observed limits at the 95\% confidence-level on $\sigma(p\overline{p}\to WH)\times\textrm{B}H\to b\overline{b}$ 
  divided by the Standard Model expectation, as a function of the Higgs mass, for the D\O\ NN+ME approach and the CDF NEAT combination.}
\begin{tabular}{c|ccccccccccc}
\hline\hline
Mass $m_H$ (GeV) & 100 & 105 & 110 & 115 & 120 & 125 & 130 & 135 & 140 & 145 & 150 \\
\hline 
Expected (D\O\ ) & 5.3 & 4.9 & 5.8 & 6.4 & 7.5 & 9.5 & 13.7 & 16.1 & 23.0 & 36.1 & 56.0 \\
Observed (D\O\ ) & 5.2 & 4.2 & 5.1 & 6.7 & 8.2 & 9.8 & 16.7 & 17.3 & 23.3 & 43.7 & 52.4 \\
Expected (CDF)   & 3.5 & 3.8 & 4.1 & 4.8 & 5.9 & 7.2 & 8.7 & 12.2 & 17.5 & 25.6 & 40.5 \\   
Observed (CDF)   & 3.3 & 3.6 & 4.9 & 5.6 & 5.9 & 8.0 & 8.9 & 13.2 & 26.5 & 42.2 & 75.5 \\ 
\hline\hline
\end{tabular}
\label{tab:limit_table}
\end{center}
\end{table*}

\bigskip 

\end{document}